\documentclass[prd,aps,10pt,showkeys,nofootinbib,showpacs,twocolumn]{revtex4-1}

\usepackage{amsmath,amssymb,amsfonts,amsthm}
\usepackage{amsbsy} 
\usepackage{epsfig}
\usepackage{latexsym}
\input amssym.def
\input amssym.tex

\usepackage{hyperref}

\newcommand{\oarX}[1]{\href{http://arxiv.org/abs/#1}{{\ttfamily #1}}}
\newcommand{\arX}[1]{\href{http://arxiv.org/abs/#1}{{\ttfamily arXiv:#1}}}

\def\barr{\begin{array}}
\def\earr{\end{array}}
\def\half{\frac{1}{2}}
\def\ben{\begin{equation}}
\def\een{\end{equation}}
\def\bs{\begin{subequations}}
\def\es{\end{subequations}}
\def\bena{\begin{eqnarray}}
\def\eena{\end{eqnarray}}

\def\bR{\mathbb{R}}

\def\SU{{\rm SU}}

\def\im{{\rm i}}

\def\be{\begin{equation}}
\def\ee{\end{equation}}

\def\bes{\begin{eqnarray}}
\def\ees{\end{eqnarray}}

\newcommand{\dd}{\mathrm{d}}

\begin{document}

\title{Cosmological perturbations from full quantum gravity}

\author{Steffen Gielen}
\affiliation{Perimeter Institute for Theoretical Physics, 31 Caroline St. N., Waterloo, Ontario N2L 2Y5, Canada}
\affiliation{Canadian Institute for Theoretical Astrophysics (CITA), 60 St.~George Street, Toronto, Ontario M5S 3H8, Canada}
\affiliation{School of Mathematical Sciences, University of Nottingham, University Park, Nottingham NG7 2RD, UK}
\email{steffen.gielen@nottingham.ac.uk}
\author{Daniele Oriti}
\affiliation{Max Planck Institute for Gravitational Physics (Albert Einstein Institute), Am M\"uhlenberg 1, D-14476 Potsdam-Golm, Germany, EU}
\email{daniele.oriti@aei.mpg.de}

\date{\today}


\begin{abstract}
The early universe provides an opportunity for quantum gravity to connect to observation by explaining the large-scale structure of the Universe. In the group field theory (GFT) approach, a macroscopic universe is described as a GFT condensate; this idea has already been shown to reproduce a semiclassical large universe under generic conditions, and to replace the cosmological singularity by a quantum bounce. Here we extend the GFT formalism by introducing additional scalar degrees of freedom that can be used as a physical reference frame for space and time. This allows, for the first time, the extraction of correlation functions of inhomogeneities in GFT condensates: in a way conceptually similar to inflation, but within a quantum field theory of both geometry and matter, quantum fluctuations of a homogeneous background geometry become the seeds of cosmological inhomogeneities. We find approximately scale-invariant initial quantum fluctuations in the local volume, with naturally small amplitude; this behaviour extends to other quantities such as the matter density. These results confirm the potential of GFT condensate cosmology to provide a purely quantum gravitational foundation for the understanding of the early universe.
\end{abstract}


\maketitle

\section{Introduction}
Cosmology provides the most promising avenue for connecting quantum gravity to observable physics; this has motivated much work in particular on models replacing the big bang with a bounce \cite{bouncers}. Since our Universe is very well described on large scales by a simple Friedmann-Lema\^{i}tre-Robertson-Walker (FLRW) metric with linear perturbations, one then looks for a manageable approximation or truncation of quantum gravity to nearly homogeneous and isotropic universes. 

In the last years, a new promising approach has emerged. In the group field theory (GFT) formalism for quantum gravity \cite{GFT} (itself a second quantized formalism for loop quantum gravity (LQG) \cite{LQG2nd} and an enrichment of random tensor models \cite{tensors} by group theoretic data), in which space and time are fundamentally made up of discrete ``atoms of geometry,'' one can describe a macroscopic, homogeneous universe as a {\em condensate}, a highly coherent configuration of many such atoms. Condensates realize a natural quantum notion of homogeneity---the condensation of many quanta into a single microscopic quantum state---and the idea that spacetime could be a type of Bose-Einstein condensate had been considered earlier \cite{becproposals}. In GFT, such condensates describe spatially homogeneous universes \cite{GFTcond}. By coupling to a massless scalar (clock) field, it was shown such universes satisfy the Friedmann dynamics of classical general relativity (GR) in a semiclassical regime \cite{GFTFRW}; the semiclassical regime is reached for generic initial conditions \cite{genericcond}. In addition, at high curvatures such condensates undergo a bounce similar to the one seen in loop quantum cosmology. For some GFT models, this bounce can be followed by a long lasting phase of acceleration, without the need to introduce an inflaton \cite{GFTinteract, marcomairi}. 

An open question in the study of GFT condensates (as in other approaches deriving cosmology from full quantum gravity) has been to extend these results from exactly homogeneous to inhomogeneous universes, i.e., to realistic and testable situations. In previous studies \cite{GFTperturb1,GFTperturb2}, a major obstacle was the localization of perturbations in a fully background-independent context, without a manifold or coordinates. Ideas from quantum cosmology such as a Born-Oppenheimer approximation for inhomogeneities \cite{qcinhomo} are not directly applicable to GFT where no separation of perturbation modes, e.g., as eigenmodes of a Laplacian, is readily available. 

Our starting point is to realize that the related problem of  ``localizing events in time'' was solved by introducing a scalar field, used as a ``clock'' to label evolution of the geometry; the problem of ``localizing events in spacetime'' is then solved by coupling {\em four} scalar fields (in four spacetime dimensions) to gravity, using these scalars as relational clocks and rods, i.e., as a physical coordinate system. This idea has a long history in classical and quantum gravity, the most famous example perhaps being Brown-Kucha\v{r} dust \cite{brownk}. Such models, in which one can solve the constraints of canonical GR and define observables on a physical phase space, have had numerous applications in LQG \cite{refLQG}.

We define a class of GFT models for gravity coupled to four reference scalar fields $\phi^I$, $I=0,\ldots,3$, generalizing Ref.~\cite{GFTFRW}. 
This allows us to define observables that correspond to a local volume element at each point in spacetime, and hence capture (scalar) inhomogeneities. 

Working in the mean-field approximation to the full quantum GFT, the effective dynamics for geometric observables can be extracted by the same methods as in Ref.~\cite{GFTFRW}. We then assume a background continuum geometry that is homogeneous, corresponding to a condensate state independent of the ``rod'' fields, to reproduce the usual setup for cosmological perturbations.

Next we compute quantum fluctuations of local volume observables in such a quantum state, staying within the full quantum gravity framework, but in a hydrodynamic approximation. We propose their two-point function as the relevant quantity in order to compare to standard cosmology and observation, and show that this is nonvanishing for a homogeneous condensate, very similar to how inhomogeneities arise from quantum vacuum fluctuations in inflation. The results outline a concrete, workable formalism for deriving a power spectrum of cosmological perturbations directly from a theory of quantum gravity, and bring quantum gravity closer to observational tests.

\section{Relational clocks and rods}

We introduce physical reference frames and define relational dynamics first in classical GR, to later implement these ideas in the quantum GFT formalism. Reference matter backreacts on the geometry, although we will consider limiting cases in which this backreaction can be negligible.

First, consider a single massless, free scalar field, used as a relational clock in a flat FLRW metric, with action
\ben
S_\phi = -\half \int \dd^4 x\;\sqrt{-g}\;g^{\mu\nu}\partial_\mu\phi\partial_\nu\phi = V_0 \int \dd t\;\frac{a^3}{2N}\dot\phi^2
\label{clockaction}
\een
where $V(t)=a^3(t) V_0$ is the 3-volume of space given in terms of a fiducial volume $V_0$. The conjugate momentum $\pi_\phi:=V\dot\phi/N$ is a conserved quantity; for any choice of time variable $t$, then, $\phi(t)$ is strictly monotonic (unless $\pi_\phi=0$, which has to be excluded). Hence $\phi$ is a good clock, and the dynamics of the Universe can be expressed in terms of $\phi$; the Friedmann equation becomes
\ben
\left(\frac{1}{V}\frac{\dd V}{\dd \phi}\right)^2 = 9 \left(\frac{\dot{a}}{a N}\frac{V}{\pi_{\phi}}\right)^2 = 12\pi G\,,
\label{relfriedmann}
\een
and its solutions are
\ben
V(\phi)=\alpha\exp(\pm\sqrt{12\pi G}\phi)\,,
\label{friedmannvol}
\een
where the sign depends on the choice of time orientation. 

The scalar action (\ref{clockaction}) is invariant under translations $\phi(t)\mapsto \phi(t)+\phi_0$ and time reversal $\phi(t)\mapsto -\phi(t)$.

This construction can be straightforwardly generalized: given four scalars $\phi^I$, one can identify the points $\{p\}$ of an open (connected) region by the values $\phi^I(p)$ if the gradients of $\phi^I$ are everywhere nondegenerate, $\det(\partial_\alpha\phi^I)\neq 0$. 

Similarly to a clock scalar field, one has to impose symmetries on the dynamics of these four scalars to be used as a material reference frame. For instance, consider the class of models in Ref.~\cite{refLQG}, with action
\bena
S_{{\rm m}}&=&-\half\int \dd^4 x\;\sqrt{-g}\left(g^{\mu\nu}\left[\rho\,\partial_\mu T\,\partial_\nu T + A(\rho)V_\mu\,V_\nu\right.\right.\nonumber
\\&&\left.\left.+2B(\rho)\partial_\mu T\,V_\nu\right]+\Lambda(\rho)\right)\,,\quad V_\mu:=W_k\partial_\mu Z^k
\label{generalaction}
\eena
depending on eight scalars $(T,Z^k,\rho,W_k)$. The dynamical fields $T$ and $Z^k$ give a local reference frame for space and time. Depending on $A(\rho),\; B(\rho)$ and $\Lambda(\rho)$, Eq.~(\ref{generalaction}) can reduce to Brown--Kucha\v{r} dust or null, nonrotational or Gaussian dust. Equation (\ref{generalaction}) is invariant under constant shifts in $T$ and $Z^j$, sign reversal of all four fields,
\bena
T(x)\mapsto T(x)+T_0\,,\quad Z^j(x)\mapsto Z^j(x)+Z_0^j\,,
\\(T(x)\mapsto -T(x)\,,\;Z^j(x)\mapsto -Z^j(x))\,,
\label{symmetries}
\eena
and ${\rm O}(3)$ transformations
\ben
Z^k(x)\mapsto {O^k}_j\,Z^j(x)\,,\quad W_k(x)\mapsto {(O^{-1})_k}^j W_j(x)\,.
\label{rotation}
\een
Equation (\ref{rotation}) implements isotropy of space: rotating the ``rods'' will define another, equivalent set of rods. We will assume all these transformations are symmetries of our reference scalar matter; they would also be symmetries of the coordinates of a good reference frame.

\section{Group field theory with reference scalar matter}

We work with GFT models for gravity coupled to four reference scalar fields, defined in analogy to known models for gravity and a single (clock) field \cite{GFTFRW,lioritizhang}. 

The basic ingredient of any GFT is a quantum field on an abstract group manifold \cite{GFT}, whose excitations form quanta of geometry labeled by data in the domain space of this field. We picture these quanta as tetrahedra equipped with a discrete ${\rm SU}(2)$ connection (parallel transports across the four faces) and with real labels for the scalar degrees of freedom. The same variables are associated to an LQG spin network vertex with four open links, for gravity coupled to four scalar fields \cite{LQG2nd}. Concretely, our GFT field is a complex field on $\SU(2)^4\times\bR^4$, denoted by $\varphi(g_I,\phi^J)$ where $g_I\in\SU(2), \phi^J\in\bR$.

One can then define the quantum GFT in the path integral or operator formalism; the latter is well suited for the study of GFT condensates \cite{LQG2nd,GFTcond}. Here one postulates canonical commutation relations 
\ben
\left[\hat\varphi(g_I,\phi^J),\hat\varphi^\dagger(g'_I,\phi'^J)\right]=\int \dd h \,\delta^4(g'_Ihg^{-1}_I)\delta^4(\phi^J-\phi'^J)
\een
while two $\hat\varphi$ or two $\hat\varphi^\dagger$ operators commute.

The Hilbert space is defined starting from a ``no-space'' vacuum $|\emptyset\rangle$, annihilated by $\hat\varphi(g_I,\phi^J)$. The bosonic excitations over $|\emptyset\rangle$, created by $\hat\varphi^\dagger(g_I,\phi^J)$, are interpreted as geometric tetrahedra. A state describing a macroscopic, approximate continuum geometry contains a very large number (potentially infinite) of such excitations.

The dynamics is governed by an action of the form
\ben
S[\varphi,\bar\varphi]=-\int \dd^4 g\,\dd^4\phi\;\bar\varphi(g_I,\phi^J)\mathcal{K}\varphi(g_I,\phi^J) + \mathcal{V}[\varphi,\bar\varphi]
\label{gftaction}
\een
where the kernel $\mathcal{K}$ is taken to be local and contain derivatives with respect to $g_I$ and $\phi^J$. The precise forms of $\mathcal{K}$ and $\mathcal{V}$ will not be used in the following. They can be chosen such that the GFT Feynman amplitudes correspond to the amplitudes of a given spin foam model \cite{SFrev}, i.e., to a discrete path integral for gravity coupled to four scalar fields. The perturbative expansion in Feynman diagrams is then a sum over such path integrals for different discretizations \cite{GFT, LQG2nd}. In order to have a sum over simplicial lattices, the interaction $\mathcal{V}$ would involve five fields, gluing five tetrahedra to a 4-simplex. Other interactions, and general forms of $\mathcal{K}$, are suggested by work on random tensor models and GFT renormalization.

We are interested in models that use scalar fields as reference matter. Following the above discussion, we assume that the GFT dynamics is invariant under
\begin{itemize}
\item[(i)] arbitrary (constant) shifts in $\phi^I$,
\item[(ii)] the parity/time-reversal transformation $\phi^I\mapsto -\phi^I$,
\item[(iii)] rotations $\phi^i\mapsto {O^i}_j\phi^j$ where $i,j=1,2,3$.
\end{itemize}
The first of these forbids explicit dependence on $\phi^I$.

We then work in an effective field theory/hydrodynamic expansion of $\mathcal{K}$ in derivatives with respect to the $\phi^J$ (as developed in Refs.~\cite{GFTFRW,lioritizhang}); this leads to an effective low-energy GFT dynamics that can be compared with cosmology on large scales, where one can truncate $\mathcal{K}$ to second derivatives. 

The assumed symmetries (i)--(iii) force this derivative expansion to be of the form
\ben
\mathcal{K}=\mathcal{K}^0+\mathcal{K}^1\Delta_{\phi^i}+\tilde{\mathcal{K}}^1\partial^2_{\phi^0}+\ldots\,,\quad \Delta_{\phi^i}\equiv\sum_{i=1}^3\partial^2_{\phi^i}
\label{kinetic2}
\een
where $\ldots$ includes fourth and higher derivatives.

Beyond the symmetries (i)--(iii), we make no assumptions about the form of $\mathcal{V}$. We will employ a weak-coupling approximation in which the effect of $\mathcal{V}$ on the dynamics is negligible. 

\section{Effective cosmological dynamics}

The proposal of GFT condensate cosmology \cite{GFTcond} is that a macroscopic, nearly homogeneous universe is well approximated by a {\em GFT condensate} with a nonvanishing field expectation value, $\sigma(g_I,\phi^J):=\langle\hat\varphi(g_I,\phi^J)\rangle \neq 0$.
In the {\em mean-field approximation}, this condition is implemented by choosing the coherent state
\ben
|\sigma\rangle \equiv N(\sigma)\exp\left(\int \dd^4 g\,\dd^4\phi\;\sigma(g_I,\phi^J)\hat\varphi^\dagger(g_I,\phi^J)\right)|\emptyset\rangle
\label{state}
\een
and all dynamical information is determined by the mean field $\sigma(g_I,\phi^J)$. This corresponds to the Gross-Pitaevskii approximation for weakly interacting Bose-Einstein condensates \cite{BEClit}. One then considers the expectation value 
\bena
0&=&\langle\sigma|\frac{\delta S[\varphi,\bar\varphi]}{\bar\varphi(g_I,\phi^J)}|\sigma\rangle = \frac{\delta S[\sigma,\bar\sigma]}{\bar\sigma(g_I,\phi^J)}
\\&=&\left(\mathcal{K}^0+\mathcal{K}^1\Delta_{\phi^i}+\tilde{\mathcal{K}}^1\partial^2_{\phi^0}+\ldots\right)\sigma(g_I,\phi^J)-\frac{\delta \mathcal{V}[\sigma,\bar\sigma]}{\bar\sigma(g_I,\phi^J)}\,, \nonumber
\eena
the GFT analogue of the Gross-Pitaevskii equation for a Bose-Einstein condensate. 

We neglect higher than second derivatives, and use an approximation in which the contribution of $\mathcal{V}$ is neglected. The latter is compatible with the weak correlations in the simple state (\ref{state}) and, tentatively, with small spatial gradients of the effective geometry. Including interactions is possible \cite{GFTinteract}, but we will show that our approximation already allows for interesting cosmological dynamics. 

The GFT equation of motion for $\sigma$ becomes
\ben
\left(\mathcal{K}^0+\mathcal{K}^1\Delta_{\phi^i}+\tilde{\mathcal{K}}^1\partial^2_{\phi^0}\right)\sigma(g_I,\phi^J)=0\,.
\een

We further restrict $\sigma$ to isotropic (equilateral) tetrahedra \cite{GFTFRW} (again this can be relaxed \cite{anisotr}); $\sigma$ can then be expanded in irreducible $\SU(2)$ representations as
\ben
\sigma(g_I,\phi^J) = \sum_{j=0}^\infty \sigma_j(\phi^J) \,{\bf D}^j(g_I)
\label{peterweyl}
\een
where the ${\bf D}^j(g_I)$ encode the equilateral shape of the tetrahedra. Because this shape is taken to be fixed, $\sigma$ only depends on a single $j$, whose value specifies the local volume and thus the cosmological scale factor. The volume can be computed within full GFT as the expectation value of a second quantized operator, see below. 

The isotropic mean field $\sigma_j(\phi^J)$ then satisfies
\ben
\left(-B_j+A_j\partial^2_{\phi^0}+C_j\,\Delta_{\phi^i}\right) \sigma_j(\phi^J)=0\,;
\label{effdyn}
\een
$\mathcal{K}^0$, $\mathcal{K}^1$ and $\tilde{\mathcal{K}}^1$ have been rewritten as $j$-dependent couplings with no further derivatives.

The results of Ref.~\cite{GFTFRW} are recovered for a mean field of the form
\ben
\sigma_j(\phi^J)\equiv\sigma_j^{0}(\phi^0)\,,
\label{meanfield}
\een
with a relational 3-volume operator at ``time'' $\phi^0$
\ben
\hat{V}(\phi^0)=\int \dd^4 g\,\dd^4 g'\;\hat\varphi^\dagger(g_I,\phi^0)V(g_I,g'_I)\hat\varphi(g'_I,\phi^0)\,.
\label{volumeop}
\een
$V(g_I,g'_I)$ are matrix elements of the LQG volume operator \cite{LQGvolume} between single-vertex spin network states.

Given a GFT state, $\langle\hat{V}(\phi^0)\rangle$ gives its total 3-volume at relational time $\phi^0$. This appears in the Friedmann equation (\ref{relfriedmann}), which connects GFT condensates to cosmology.

In this case, generic initial conditions lead to a semiclassical regime, in which the Universe expands to macroscopic size \cite{GFTFRW,genericcond} and the 3-volume follows the classical Friedmann solution (\ref{friedmannvol}). At small volumes, the Universe undergoes a bounce, avoiding the classical singularity \cite{GFTFRW}.

For example, if only a single spin $j_0$ is excited, the 3-volume behaves as  
\ben
\langle\hat{V}(\phi^0)\rangle\stackrel{\phi^0\rightarrow\pm\infty}{\sim} |\sigma^{\pm}|^2 \exp\left(\pm 2\sqrt{\frac{B_{j_0}}{A_{j_0}}}\phi^0\right)
\label{volumesolution}
\een
for generic initial conditions ($\sigma^\pm\neq 0$), if $B_{j_0}/A_{j_0}>0$; this is precisely Eq.~(\ref{friedmannvol}) with $B_{j_0}/A_{j_0}=:3\pi G$. $V(\phi)$ interpolates between the classical contracting and expanding solutions, and only ever vanishes for special initial conditions \cite{GFTFRW,genericcond,marcomairi}. Including interactions can prolong the accelerated expansion after the bounce and cause a later recollapse, producing a cyclic cosmology \cite{GFTinteract}.

\section{Volume perturbations in GFT condensates}

Our GFT model has enough degrees of freedom to describe inhomogeneous quantum geometries and their dynamics. Here we consider situations relevant for fundamental cosmology: we study quantum fluctuations of the local 3-volume around a nearly homogeneous background, seeking a quantum gravitational mechanism for explaining the origin of inhomogeneities, in a similar spirit to the inflationary paradigm, where this mechanism is the imprint of quantum fluctuations of the inflaton \cite{mukhanov}. We show how such mechanism, natural in any quantum field theory for gravity and matter, is realized by GFT condensates, without requiring an inflaton.

We start by generalizing Eq.~(\ref{volumeop}) to a GFT for gravity coupled to four reference scalar fields $\phi^I$,
\ben
\hat{V}(\phi^J)=\int \dd^4 g\,\dd^4 g'\;\hat\varphi^\dagger(g_I,\phi^J)V(g_I,g'_I)\hat\varphi(g'_I,\phi^J)\,.
\label{volumeelement}
\een
Now all four $\phi^J$ take fixed values: $\hat{V}(\phi^J)$ defines a local volume element at the spacetime point specified by values of the reference fields. The total 3-volume (\ref{volumeop}) is obtained by integrating over the rods $\phi^i$,
\ben
\hat{V}(\phi^0)\equiv \int \dd^3 \phi\; \hat{V}(\phi^0,\phi^i)\,.
\label{rodintegral}
\een

In a simple coherent state of the form (\ref{state}), the expectation value of $\hat{V}(\phi^J)$ can be evaluated immediately,
\ben
\langle\hat{V}(\phi^J)\rangle = \int \dd^4 g\,\dd^4 g'\;\bar\sigma(g_I,\phi^J)V(g_I,g'_I)\sigma(g'_I,\phi^J)\,.
\een
For the isotropic wave function (\ref{meanfield}), we obtain
\ben
\langle\hat{V}(\phi^J)\rangle = \sum_{j=0}^\infty V_j |\sigma_j^0(\phi^0)|^2\,,
\een
with eigenvalues $V_j\sim V_{{\rm Pl}}\, j^{3/2}$ of the volume operator. The local and total 3-volume coincide [up to regularization of the integral over $\phi^i$ in Eq.~(\ref{rodintegral})], as expected in a homogeneous geometry. 

In cosmology the pattern of cosmic structure is encoded in correlation functions for geometric observables. Here we focus on local volume fluctuations $\langle \hat{V}(\phi^J) \hat{V}(\phi'^J)\rangle$ in the state (\ref{state}), which depend on the one-body matrix elements $V^2(g_I,g'_I)$ of the {\em squared} volume operator. Using ``squared matrix elements'' to characterize perturbations has been suggested before \cite{GFTperturb2}, but without rods only global information was obtained. Here we can extract local information about cosmological perturbations: Fourier transforming from $\phi^i$ to their momenta $k_i$ introduces a notion of wave number, defined with respect to the reference matter. 

We then obtain, within the full quantum gravity formalism, a power spectrum of cosmological perturbations. Consider a mean field perturbed around homogeneity,
\ben
\sigma_j(\phi^J)= \sigma_j^0(\phi^0)(1+\epsilon\,\psi_j(\phi^J))\,.
\label{perturbedmeanfield}
\een
In this state, fluctuations of the volume take the form
\bena
&&\langle \hat{\tilde{V}}(\phi^0,k_i) \hat{\tilde{V}}(\phi'^0,k'_i)\rangle - \langle \hat{\tilde{V}}(\phi^0,k_i)\rangle\langle \hat{\tilde{V}}(\phi'^0,k'_i)\rangle \nonumber
\\&=&\; \delta(\phi^0-\phi'^0) \sum_j V_j^2\,|\sigma_j^0(\phi^0)|^2\bigl[(2\pi)^3\delta^3(k_i+k'_i)\nonumber
\\&&\;+ \,\epsilon\,\left(\tilde\psi_j(\phi^0,k_i+k'_i)+\overline{\tilde\psi_j(\phi^0,-k_i-k'_i)}\right)\bigr]\,, \label{powersp}
\eena 
where we have Fourier transformed $\hat{V}$ and $\psi_j$; the delta function in $\phi^0$ arises because $\hat{V}(\phi^J)$ is a density on scalar field space. This power spectrum is a genuine quantum correlation in the GFT condensate. 

Remarkably, the dominant part of the power spectrum
\bena
(2\pi)^3\delta^3(k_i+k'_i)\delta(\phi^0-\phi'^0) \sum_j V_j^2\,|\sigma_j^0(\phi^0)|^2
\eena
is naturally scale invariant: it only depends on $\phi^0$. This property follows from computing cosmological perturbations on an exactly homogeneous background. Due to quantum fluctuations, even in this case Eq.~(\ref{powersp}) is not zero: it must then be scale invariant, with scale defined by the reference matter. Within our mean-field approximation, scale invariance and translational invariance, as expressed by the momentum delta function in Eq.~(\ref{powersp}), are necessarily connected. 

In cosmology, the usual notion of scale invariance refers to the dimensionless power spectrum, which is not directly the quantity we compute here. Converting our expressions into those appearing in the measured spectrum of inhomogeneities may introduce a dependence on $k$, in particular since our notion of scale refers to reference matter, not Cartesian coordinates. This is why we do not take over the usual terminology from cosmology, but focus on the spectrum that can be computed.
 
Deviations from exact scale invariance are encoded in the last line of Eq.~(\ref{powersp}). They arise from inhomogeneous fluctuations around a homogeneous condensate, which should generically be present; approximate scale invariance is intrinsically linked to such GFT fluctuations being small. These fluctuations must solve the mean-field condensate dynamics, so both their exact shape and their relative amplitude are determined dynamically. Further deviations will come from relaxing the mean-field approximation, i.e., from using more refined quantum states. Such deviations from scale invariance depend both on the coupling of inhomogeneities with the homogeneous background and on their own dynamics, as expected physically and in agreement with usual cosmological perturbations. They are fully determined by the GFT perturbation density field, itself a solution to mean-field equations. A more detailed study of solutions of such perturbed equations, and their initial conditions, would be crucial to identify the precise form of these deviations. 

The amplitude of volume fluctuations relative to the background, i.e., of $\widehat{\delta  \tilde{V}}(\phi^0,k_i)\equiv\hat{\tilde{V}}(\phi^0,k_i)/\langle \hat{V}(\phi^0)\rangle$,  is obtained by dividing Eq.~(\ref{powersp}) by the squared background volume $\langle \hat{V}(\phi^0)\rangle^2 \equiv (\int \dd\phi^i \sum_j V_j |\sigma_j^0(\phi^0)|^2)^2$. This amplitude is of order $1/N$, for $N\gg 1$ quanta in the condensate. 
For instance, considering only the scale invariant contribution and with only a single spin $j_0$ excited, the power spectrum of such perturbations is
\ben
\mathcal{P}_{\delta V}(k)=\frac{V_{j_0}^2\,|\sigma_{j_0}^0(\phi^0)|^2}{(\int \dd\phi^i\; V_{j_0} |\sigma_{j_0}^0(\phi^0)|^2)^2}=\frac{V_{j_0}}{(\int \dd\phi^i)V(\phi^0)}\,,
\een
with $V(\phi^0)=N(\phi^0)V_{j_0}$. A small amplitude of scalar perturbations, decreasing as the Universe expands, arises naturally from the simplest GFT condensates.

For $C_j/B_j<0$ in Eq.~(\ref{effdyn}), inhomogeneous perturbations decay relative to the homogeneous background at large volumes; one approaches scale invariance even more closely, further suppressing the deviations coming from the inhomogeneous term. GFT interactions that produce a long-lasting accelerated expansion after the bounce \cite{GFTinteract} further suppress deviations from scale invariance. 

The choice of vacuum, e.g., as made in inflation, is replaced by the GFT condensate state (\ref{state}) that refers to both quantum geometric and matter degrees of freedom. This is because such fluctuations are computed directly within the complete quantum gravity formalism, which also defines the ultraviolet completion of the theory.  

\section{Extending the formalism to density perturbations}

We showed that the statistics of scalar perturbations can be computed explicitly for GFT condensates, fully within the quantum gravity formalism, assuming that the mean field describing the condensate is close to homogeneity. We found a nearly scale-invariant power spectrum for volume perturbations, with naturally small amplitude. Exact scale invariance is found if the mean field is exactly homogeneous.

The reason for using volume perturbations was that these are simplest to compute in our formalism: they can be expressed through expectation values and fluctuations of local volume elements, given by the GFT Fock space operator (\ref{volumeelement}). Our principal goal was to show the feasibility of these computations in the full quantum gravity formalism, and the generic features of the results. 

Connecting our results to observation, however, ultimately requires also considering perturbations in the matter density. Their relation to volume perturbations is in general gauge dependent, so one cannot {\em a priori} assume that the spectrum of volume perturbations we found can be translated into an observational prediction. To close this gap, in this section we show how to extend the arguments to perturbations in the matter density. This is a more involved calculation since in our formalism, in which spacetime is a many-body quantum system, the natural observables are ``extensive'' quantities such as volume or total energy. The matter energy density is obtained from taking a quotient of expectation values of primary extensive quantities.

We start with the kinetic energy density in a scalar field, which is classically given by $\rho_{{\rm kin}}^I = (\pi_\phi^I)^2/(2 V^2)$ where $V\equiv\sqrt{h}$ corresponds to the local volume element, and $\pi_\phi^I$ is the momentum conjugate to the scalar field $\phi^I$. Hence, to construct a kinetic energy density we need to consider operators corresponding to the conjugate momenta for the four scalars; these are 
\bena
\hat{\pi}_\phi^I(\phi^J)&=&-\frac{\im}{2} \int \dd^4 g\left(\hat\varphi^\dagger(g_I,\phi^J)\frac{\partial\hat\varphi(g_I,\phi^J)}{\partial\phi^I}\right.\nonumber
\\&&\left.-\frac{\partial\hat\varphi^\dagger(g_I,\phi^J)}{\partial\phi^I}\hat\varphi(g_I,\phi^J)\right)\,,
\label{scalarmomentum}
\eena
as already defined for the homogeneous case in previous work \cite{GFTFRW}, where the scalar field momentum enters correctly in the energy density appearing in the Friedmann equations.

From the expectation values of (\ref{volumeelement}) and (\ref{scalarmomentum}), we can then define
\ben
\rho_{{\rm kin}}^I(\phi^J) = \frac{1}{2}\left(\frac{\langle \hat{\pi}_\phi^I(\phi^J)\rangle}{\langle\hat{V}(\phi^J)\rangle}\right)^2\,,
\een
and the total kinetic energy is $\rho_{\rm kin}=\sum_I \rho_{\rm kin}^I$. At leading order, fluctuations in the kinetic energy density are given by $\delta\rho/\rho = \sum_I\pi_\phi^I\delta \pi_\phi^I/(\rho V^2)-2\delta V/V$. Their two-point function is
\bena
&&\frac{\langle \delta\rho_{{\rm kin}}(\phi^0,k_i)\delta\rho_{{\rm kin}}(\phi'^0,k'_i)\rangle}{\rho_{{\rm k}}(\phi^0)^2}\nonumber
\\&=&\frac{\sum_{IJ}\pi_\phi^I(\phi^0)\pi_\phi^J(\phi^0)\langle\delta\pi_\phi^I(\phi^0,k_i)\delta\pi_\phi^J(\phi'^0,k'_i)\rangle}{\rho_{{\rm k}}(\phi^0)^2\,V(\phi^0)^4}\nonumber
\\&&-4\frac{\sum_I\pi_\phi^I(\phi^0)\langle\delta\pi_\phi^I(\phi^0,k_i)\delta V(\phi'^0,k'_i)\rangle}{\rho_{{\rm k}}(\phi^0)\,V(\phi^0)^3}\nonumber
\\&&+4\frac{\langle \delta V(\phi^0,k_i)\delta V(\phi'^0,k'_i)\rangle}{V(\phi^0)^2}\,.
\eena
We can now simplify calculations for the right-hand side by again using a homogeneous mean field of the form
\ben
\sigma_j(\phi^J)\equiv  \sigma_j^0(\phi^0)\,.
\label{homo}
\een
For this choice of mean field, we have already computed the last term on the right-hand side and shown that it gives a scale-invariant power spectrum with small amplitude. For the other two terms, we use the fact that derivatives of $\sigma_j$ with respect to the rod fields vanish, and that we hence have $\pi_\phi^i\equiv \langle\pi_\phi^i\rangle = 0$ ($i=1,2,3$) and thus $\rho_{\rm kin}=\rho_{\rm kin}^0$. Strictly speaking, there needs to be a nonzero energy density in these fields for them to form a good reference frame. However, this energy density can be arbitrarily small, so that an infinitesimal perturbation of Eq.~(\ref{homo}) will lead to a good reference frame. We assume the validity of a perturbative expansion around Eq.~(\ref{homo}), and can consider the leading term in which such perturbations are exactly zero.
\\The fluctuations in the kinetic energy then reduce to
\bena
&&\frac{\langle \delta\rho_{{\rm kin}}(\phi^0,k_i)\delta\rho_{{\rm kin}}(\phi'^0,k'_i)\rangle}{\rho_{{\rm k}}(\phi^0)^2}\nonumber
\\&=&4\frac{\langle\delta\pi_\phi^0(\phi^0,k_i)\delta\pi_\phi^0(\phi'^0,k'_i)\rangle}{\pi_\phi^0(\phi^0)^2}-8\frac{\langle\delta\pi_\phi^0(\phi^0,k_i)\delta V(\phi'^0,k'_i)\rangle}{\pi_\phi^0(\phi^0)\,V(\phi^0)}\nonumber
\\&&+4\frac{\langle \delta V(\phi^0,k_i)\delta V(\phi'^0,k'_i)\rangle}{V(\phi^0)^2}\,.
\eena
All terms on the right-hand side now give a scale-invariant power spectrum: all expectation values involve observables that do not depend on the rod fields (neither multiplicatively or in derivatives), and the mean field does not depend on these fields either. Hence, we find a scale-invariant power spectrum even for density perturbations, with amplitude still scaling as $1/N$ (a generic property of macroscopic observables for many-particle states).

Scale invariance will be broken by two types of corrections: first, as for volume perturbations, departures from exact homogeneity in the mean field lead to non-scale-invariant terms. The details will be different for density perturbations, since the rod fields will also acquire a nonzero background energy density and hence contribute to the expressions for perturbations; for instance, we find
\ben
\langle\delta\pi_\phi^i(\phi^0,k_i)\delta\pi_\phi^j(\phi'^0,k'_i)\rangle = \frac{N}{4}k^i k^j(2\pi)^3\delta^3(k_i+k'_i)\delta(\phi^0-\phi'^0)
\een
at leading order, which breaks scale invariance.

More importantly, we have ignored gradient energy in the scalar fields in these calculations, which will be expressed in terms of more complicated observables that each involve more than one scalar field. For the contribution in gradient energy, we would in general not expect scale invariance even for a homogeneous mean field. The assumption of subdominance for the gradient energy with respect to kinetic energy is what one would expect in a near-homogeneous geometry, and it is thus reasonable in a realistic cosmological scenario. In other contexts, other terms may dominate over the kinetic energy as well, for example in slow-roll inflation; such additional terms are however not necessary, it seems, in our context. More work is certainly needed to verify whether our assumptions are dynamically justified, even in presence of more realistic matter fields.

The possibility of deviations from scale invariance for gradient energy is consistent with the fact that in classical cosmology volume and density perturbations are not necessarily proportional to one another. Indeed, any such statement depends on the chosen gauge. Consider for instance the gauge-invariant ``curvature perturbation on uniform-density hypersurfaces'' (see e.g., \cite{baumann})
\ben
-\zeta = \Psi + \frac{H}{\dot\rho}\delta\rho\,,
\een
defined in terms of a metric perturbation $\Psi$, the Hubble parameter $H$, background matter density $\rho$ and density perturbation $\delta\rho$. One can choose a gauge in which $\delta\rho=0$ and $\Psi$ is proportional to the volume perturbation, or a different gauge in which $\Psi=0$. Thus, the gauge-invariant quantity $\zeta$ can be proportional either to volume or to density perturbations, but this is not true in general. 

In our formalism, due to the introduction of reference scalar matter fields which are used as relational coordinates, there is no gauge freedom and gauge choices that make certain quantities vanish are not possible (instead, our relational coordinates define a harmonic gauge \cite{harmogauge}). We should then focus on density perturbations, and as we saw, if gradient energy contributes to density perturbations, they in general depart from the scale invariance found for volume perturbations. For subdominant gradient energy and a mean field close to homogeneity, however, we also find scale invariant density perturbations, and the general result is the same: in the regime we considered, quantum gravity naturally produces an approximately scale invariant spectrum also for density perturbations, with small amplitudes. 

The detailed relation between our gauge-invariant quantities and the perturbation variables normally used in cosmology will need to be worked out to have a full comparison with observations; we leave this to future work. This detailed comparison can be based on previous work, e.g., in the canonical gravity context \cite{Giesel:2007wi}.

\section{Discussion}

By introducing in the GFT formalism scalar field degrees of freedom that can be used as physical reference frames, we could extend the mean-field approximation for GFT condensates beyond homogeneity. This approximation has already been shown to provide an effective cosmological dynamics in which not only a semiclassical large Friedmann universe is reproduced under generic conditions, but also the cosmological singularity is replaced by a quantum bounce, followed by an accelerated phase of expansion of quantum gravity origin. We then considered the typical setup of early universe cosmology within this full quantum gravity framework: we computed the power spectrum of quantum fluctuations of the local volume in a homogeneous background geometry perturbed by small inhomogeneities. We found that this is approximately scale invariant, with a small amplitude that decreases as the volume of the Universe grows. 
This confirms the potential of the GFT condensate cosmology framework to provide a quantum gravitational foundation for early universe cosmology.

While we initially showed how volume perturbations arise as quantum fluctuations in a GFT condensate, we then also saw that similar statements can be made for density perturbations, which are more directly related to observation. In particular, for perturbations in the kinetic energy of the scalar fields, the same general conclusions follow: for the mean field (\ref{perturbedmeanfield}), quantum fluctuations of such observables have scale-invariant power spectrum at leading order, as in Eq.~(\ref{powersp}). This is in general not true for gradient (potential) energy, and hence fluctuations in the total density will in general not be scale invariant as soon as gradient energy contributes non-negligibly to the total energy density. The second main property we identified, a small amplitude scaling inversely with the particle number, is more generic and extends to other observables. 

As we have stressed, precise details of the power spectrum depend on identifying a particular choice of mean field as a solution to the condensate hydrodynamics. In addition, obtaining {\em classical} inhomogeneities from quantum fluctuations requires studying the propagation, amplification and ``freeze-out'' of the initial quantum fluctuations, which all again depend on the dynamics. The details of  this transition from the initial quantum fluctuations in the deep quantum gravity regime to classical observable inhomogeneities will be the focus of future work. It is however a remarkable result that generic properties of the spectrum of observables of cosmological interest can be identified from the general formalism of cosmology as GFT hydrodynamics alone, as we have demonstrated.

\acknowledgments
We thank E.~Wilson-Ewing for very useful comments and suggestions, and three anonymous referees for comments on previous versions that led to improvements in presentation. D.O. also acknowledges hospitality at the Perimeter Institute of Theoretical Physics, where part of this work was done. Research at Perimeter Institute is supported by the Government of Canada through the Department of Innovation, Science and Economic Development Canada and by the Province of Ontario through the Ministry of Research, Innovation and Science. The work of S.G. was supported by a Royal Society University Research Fellowship (UF160622).

\end{document}